\documentstyle[12pt,epsf]{article}
\begin{document}
\baselineskip 16pt plus 2pt minus 2pt

\newcommand{\beq}{\begin{equation}}
\newcommand{\eeq}{\end{equation}}
\newcommand{\beqa}{\begin{eqnarray}}
\newcommand{\eeqa}{\end{eqnarray}}
\newcommand{\ve}{\varepsilon}
\newcommand{\eps}{\epsilon}
\newcommand{\krig}[1]{\stackrel{\circ}{#1}}
\newcommand{\barr}[1]{\not\mathrel #1}
\def\simge{\hspace*{0.2em}\raisebox{0.5ex}{$>$}
     \hspace{-0.8em}\raisebox{-0.3em}{$\sim$}\hspace*{0.2em}}
\def\simle{\hspace*{0.2em}\raisebox{0.5ex}{$<$}
     \hspace{-0.8em}\raisebox{-0.3em}{$\sim$}\hspace*{0.2em}}

\begin{titlepage}

\hfill {INT \#DOE/ER/40561-306-INT96-00-159}\\
\hphantom{xxx} \hfill {MKPH-T-96-19}

\vspace{1.0cm}

\begin{center}
{\large { \bf Contributions of strange quarks to the
magnetic moment of the proton}}

\vspace{1.2cm}

H.-W. Hammer$^{a,b,}$\footnote{Electronic address:
hammerhw@latte.phys.washington.edu},
D. Drechsel$^{a,}$\footnote{Electronic address:
drechsel@kph.uni-mainz.de},
and
T. Mart$^{a,}$\footnote{Electronic address:
mart@kph.uni-mainz.de}

\vspace{0.8cm}

$^{a}$Universit\"at Mainz, Institut f\"ur Kernphysik, J.-J.-Becher
Weg 45\\ D--55099 Mainz, Germany\\[0.4cm]
$^{b}$Institute for Nuclear Theory, University of Washington\\ Seattle,
WA 98195, USA\\[0.4cm]
\end{center}

\vspace{1cm}

\begin{abstract}
Using the Gerasimov-Drell-Hearn sum rule and experimental total cross 
sections for photoproduction of $\eta$, $K$ and $\phi$ mesons on the proton, 
we obtain upper bounds for the contribution of strange quarks to the 
anomalous magnetic moment of the proton, $|\kappa_{s}| \le 0.20\;
\kappa_{p}$.

The proposed experiments to measure the spin-dependence of the absorption 
cross section are expected to lower these bounds considerably.
The existing data on $\eta$ production and phenomenological models
for $K$ production agree with a negative sign for $\kappa_{s}$.\\[0.3cm]
\end{abstract}

\vspace{2cm}
\vfill
\end{titlepage}

The strange vector form factors of the nucleon are an interesting 
and challenging topic in medium energy physics. They obtain 
contributions only from non-valence quarks and allow for a direct
study of the role played by strange quarks for the nucleon's response
to low- and intermediate-energy probes \cite{Mu94}.
Considerable experimental activities have been devoted 
to measure these form factors by means of 
parity violating electron scattering at MAMI, MIT/Bates, and Jefferson
Lab (formerly CEBAF) \cite{exprops}. 
At the same time, a wealth of model calculations is 
now available \cite{theorefs}. 
However, these calculations involve large theoretical uncertainties, 
because sea quark effects arise from a subtle interplay of quantum 
fluctuations which are difficult to describe. 
Recently, there has been some effort
to reduce the uncertainty in these predictions by using chiral 
perturbation theory \cite{Mu96_1} and dispersion theory in 
conjunction with unitarity \cite{Mu96_2}.
However, the aim of this letter is to point out a complementary 
way to estimate  the importance of the strange quark for the structure 
of the nucleon. 

As is well known, the ground state properties of the nucleon are connected 
with its excitation spectrum via dispersion relations and low energy theorems.
In particular, the Gerasimov-Drell-Hearn (GDH) sum rule \cite{DHG66} relates 
the anomalous magnetic moment $\kappa$ of the nucleon to the difference of 
its polarized total photoabsorption cross sections,
\beq
\label{dhg}
-\frac{\kappa^2}{4} = \frac{m^2}{8 \pi^2\alpha}
\int\limits_0^\infty \frac{d\nu}{\nu}\left(
\sigma_{1/2}(\nu) - \sigma_{3/2}(\nu) \right)=: I(Q^2=0)\; ,
\eeq
where $\sigma_{3/2}$ and $\sigma_{1/2}$ denote the cross sections for the two
possible combinations of the spins of nucleon and photon, $\alpha$ is the fine
structure constant, $\nu$ the photon energy in the laboratory frame and $m$ 
the mass of the nucleon. This sum rule is based on Lorentz and gauge 
invariance, crossing symmetry, causality and unitarity. The only further 
assumption in deriving eq. (\ref{dhg}) is the convergence of the integral. 
There are in fact very strong arguments for such a convergence from the 
Froissart bound for spin dependent processes in QCD \cite{Ba96}. A general 
review of the field can be found, e.g. in \cite{DD95}. 
In particular, we note that the sum rule can be extended to virtual
photons. The difference of the helicity cross sections is then 
related to the transverse-transverse structure function
\beq
\label{sigtt}
\sigma_{TT'} = \frac{\sigma_{3/2}-\sigma_{1/2}}{2} \; ,
\eeq
which can be measured with longitudinally polarized electrons and nucleons 
polarized in the direction of the exchanged virtual photon,
\beq
\label{dd1}
I(Q^2) = -\frac{m^2}{4\,\pi^2\alpha}\int\limits_0^\infty
\frac{d\nu}{\nu}(1-x)\;\sigma_{TT'}(\nu,Q^2)\; ,
\eeq
where $x=\frac{Q^2}{2\,m\,\nu}$ denotes the Bjorken scaling variable
and $-Q^2$ is the 4-momentum transfer of the virtual photon.
$\sigma_{TT'}$ may be expressed through a multipole decomposition.
Up to kinematical factors, one has 
\beq
\label{dd2}
\sigma_{TT'} \propto \left( -|E_{0+}|^2 -3|E_{1+}|^2 +|M_{1+}|^2
-|M_{1-}|^2 -6 E_{1+}^* M_{1+} + \ldots\; \right)\; ,
\eeq
where the dots correspond to higher multipoles \cite{DD95}.

Our aim is to use the GDH sum rule and experimental photoproduction data 
from a proton target to estimate the fraction of the magnetic moment of 
the proton  which is due to the strangeness degree of freedom. Therefore, 
we will study the contributions of photoproduction of the lightest mesons 
carrying strange quarks ($\eta, K, \phi$) to the integral on the right hand 
side of eq.(\ref{dhg}). Obviously the contributions of the various reaction 
channels add incoherently to the partial cross sections in eq. (\ref{dhg}), 
and the sum over all channels involving reaction products with strange quarks 
gives some measure of the influence of strangeness on the magnetic moment 
of the nucleon.

\begin{figure}[htb]
\epsfxsize=5in
\begin{center}
\
\epsffile{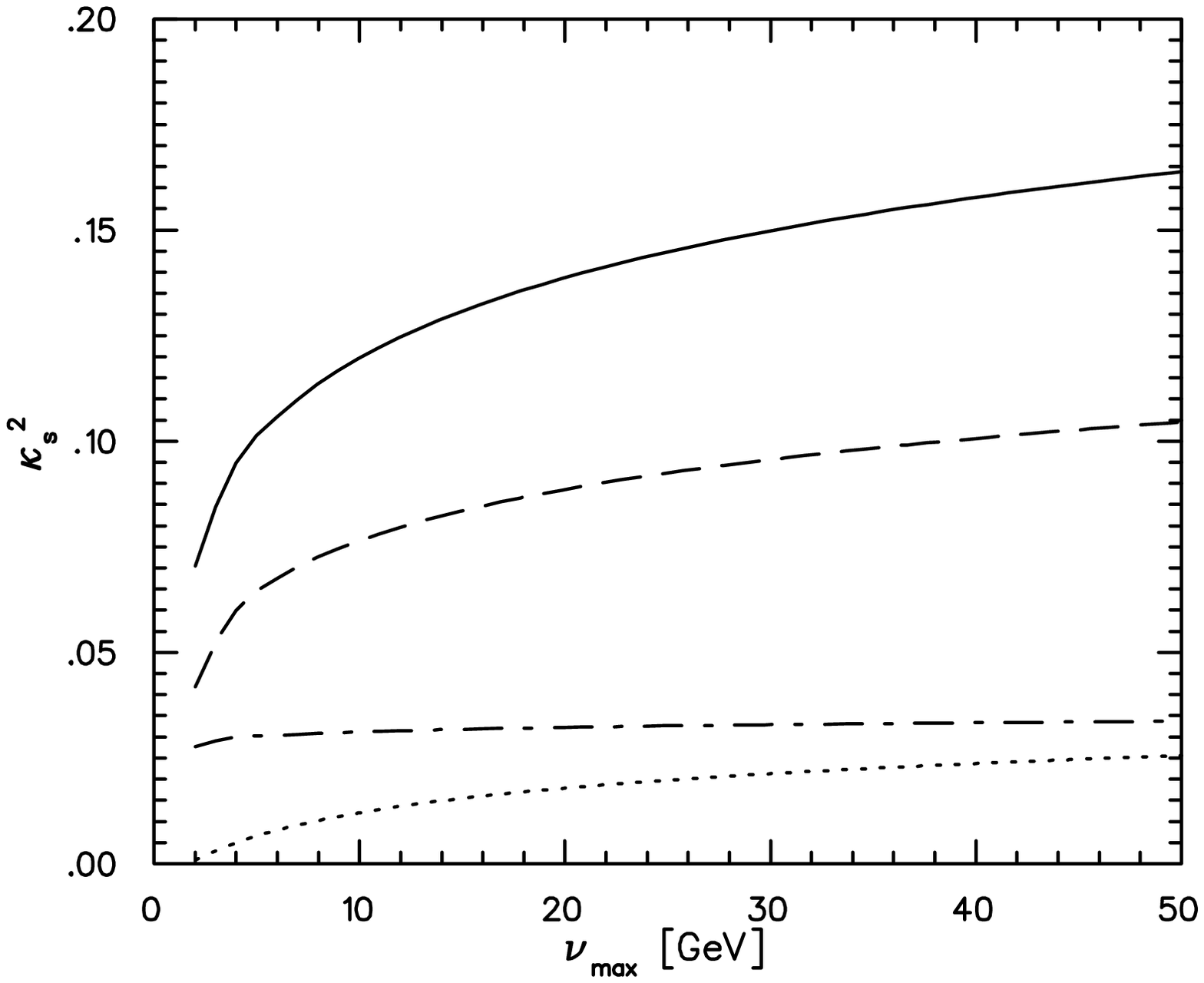}
\end{center}
\caption{\label{fig1}The dependence on $\nu_{max}$ of 
the contribution of $\phi$, $\eta$, and $K$ meson photoproduction
to the anomalous magnetic moment of the proton (upper bounds). 
The individual contributions
$\kappa_s^2(\phi)$, $\kappa_s^2(\eta)$, $\kappa_s^2(K)$, and the
sum $\kappa_s^2$ are denoted by the dotted, dash-dotted, dashed, and
solid lines, respectively.
}
\end{figure}

In phenomenological analyses \cite{Wo92}, it has been found that the GDH 
integral for the proton is already saturated for photon energies at 
$\nu_{max} \approx 2 \mbox{ GeV}$,
\beq
\label{dhg2}
\kappa_p^2 \approx \frac{m^2}{2 \pi^2\alpha}
\int\limits_0^{\nu_{max}} \frac{d\nu}{\nu}\left(
\sigma_{3/2}(\nu) - \sigma_{1/2}(\nu) \right)\; .
\eeq
Since the existing data base is not yet accurate enough to extract 
$\sigma_{3/2}$ and $\sigma_{1/2}$ for channels involving strange quarks, 
we use the total cross section as an upper bound. In the electroproduction
formalism this corresponds to using the structure function
\beq
\label{sigt}
\sigma_{T} = \frac{\sigma_{3/2}+\sigma_{1/2}}{2} \;
\eeq
instead of $\sigma_{TT'}$. With $\sigma_{3/2}$ and 
$\sigma_{1/2}$ being positive definite quantities and $\sigma_{tot}(\nu)=
2 \, \sigma_T(\nu,Q^2=0)$, we have
\beq
\label{ungl}
-\sigma_{tot}(\nu) \leq
\sigma_{3/2}(\nu) - \sigma_{1/2}(\nu)
\leq \sigma_{tot}(\nu)\;,
\eeq 
and
\beq
\label{ungldgh}
\kappa_p^2 \;\simle\, \frac{m^2}{2 \pi^2\alpha}
\int\limits_0^{\nu_{max}} \frac{d\nu}{\nu}\, \sigma_{tot}(\nu)\; .
\eeq
We will use eq. (\ref{ungldgh}) to estimate the contribution of reactions 
involving strange quarks to the anomalous magnetic moment of the proton,
and denote this contribution by $\kappa_s$. The experimental 
data for the total cross sections for photoproduction of $\eta, 
K$ and $\phi$ mesons on the nucleon are extracted from refs. 
\cite{Aa68,etadat}, \cite{Kdat} and \cite{Aa68,phidat}, respectively. 
A summary of the older data can be found in ref. \cite{LB73}.
In the low energy region the experimental data have 
been interpolated, whereas at high energies a constant cross section equal 
to the value at the highest measured energy has been assumed. 
In the case of $K$ production we have to sum over the reactions $\gamma p 
\rightarrow K^{+}\Sigma^{0}, K^{+}\Lambda$ and $K^{0}\Sigma^{+}$. 
For the $\phi$ and $\eta$ mesons which carry no net strangeness,
we explicitly consider their flavour content.
The $\phi$ is an almost pure $|\bar{s}s\rangle$ state,
since $\phi-\omega$ mixing is small ($\eps \approx 0.055$).
However, in the eta case the octet and singlet states mix with an
angle $\theta_{\eta\eta'}\approx -0.35$ leading to a flavour 
content $|\eta\rangle = 3^{-1/2}( |\bar{u}u\rangle
+|\bar{d}d\rangle -|\bar{s}s\rangle)$. Since we are only 
interested in the contribution of strange quarks to the
magnetic moment of the proton, we multiply
the $\eta$ contribution by a factor $1/3$.  
\begin{table}[htb]
\begin{center}  
\begin{tabular}{|c||c|c|c|c|c|c|} \hline
$\nu_{max}$ [GeV] & $\kappa_s^2(\phi)$ & $\kappa_s^2(\eta)$ & 
$\kappa_s^2(K)$ & $\kappa^2_s$ (sum) & $\kappa^2_s /\kappa^2_p$ [\%] 
& $|\kappa_s| /\kappa_p$ [\%]\\ \hline\hline
2.0 & 0.00 & 0.03 & 0.04 & 0.07 & 2 & 14\\
5.0 & 0.01 & 0.03 & 0.06 & 0.10 & 3 & 17\\
50.0 & 0.03 & 0.03 & 0.10 & 0.16 & 5 & 22\\
\hline
\end{tabular}
\end{center}
\caption{\label{restab}The contributions of $\phi,\eta$, and $K$
production to the anomalous magnetic moment of the proton
($\kappa_p = 1.79$), for different cut-offs $\nu_{max}$ .}
\end{table}
In Table \ref{restab} we display the results obtained with the three 
different cutoff values $\nu_{max} = 2.0, 5.0$ and $50.0 \mbox{ GeV}$. 
Since the thresholds for $\eta, K$ and $\phi$ production are considerably 
higher than the pion threshold, it is reasonable to integrate at least 
up to $\nu_{max} = 5.0 \mbox{ GeV}$.
In particular, the contribution of $\phi$ production is completely negligible 
for $\nu_{max}  = 2.0 \mbox{ GeV}$. Of course, the largest value in the table,
$\nu_{max} = 50.0 \mbox{ GeV}$, is not very realistic since at such energies 
single meson production is negligible and the approximation of a constant 
cross section certainly overestimates the effect. However, the last 
column in Table \ref{restab} shows that the final results are not very 
sensitive to the cut-off energy. The dependence on $\nu_{max}$ is displayed 
in Fig. \ref{fig1} for both the individual contributions and the sum.
The logarithmic increase
at high energies is due to the assumption of a constant cross section 
in this region. It can be noticed that even very unrealistic values of 
$\nu_{max}$ do not change the order of magnitude of the estimate. From 
Table \ref{restab} we conclude that $\kappa_{s}^{2}\le 0.04\;
\kappa_{p}^{2}$ 
is a reasonable upper bound, i.e. $|\kappa_{s}|\le 0.20\;\kappa_{p}$. 
This bound is in agreement with most of the theoretical calculations 
\cite{theorefs,Mu96_1,Mu96_2}. In our calculation, 
$K$ production is the most important reaction, whereas the 
contributions of $\phi$ and $\eta$ mesons are considerably smaller.

\begin{figure}[htb]
\epsfxsize=13cm
\begin{center}
\
\epsffile{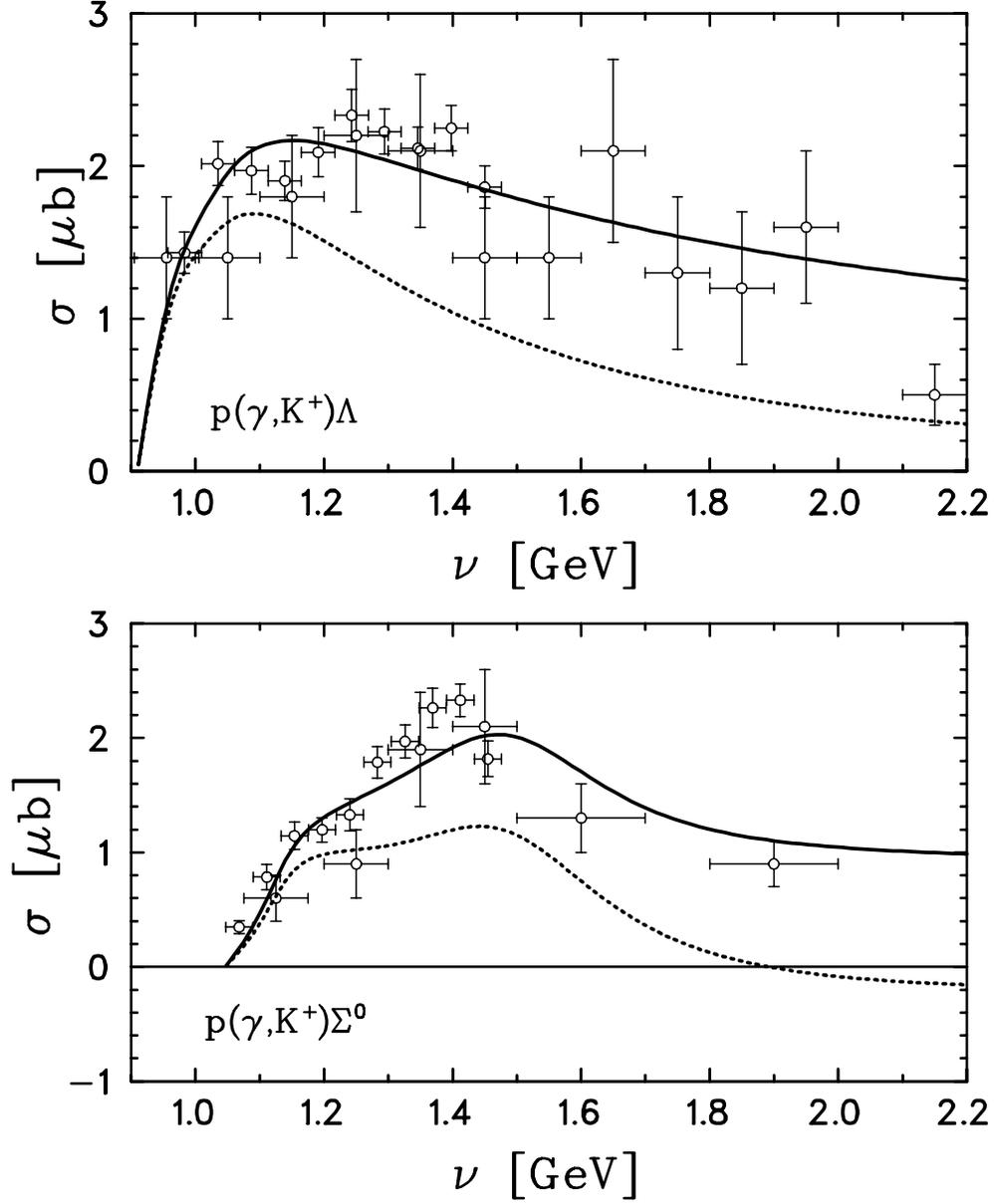}
\end{center}
\caption{\label{fig2}The response functions $\sigma_T$ (solid lines)
and $-\sigma_{TT'}$ (dotted lines) at $Q^2$=0 for the reactions $\gamma + p
\rightarrow K^+ + \Lambda$ (upper panel) and $\gamma + p \rightarrow 
K^+ + \Sigma^0$ (lower panel) calculated with the model of Mart et al.
\protect\cite{mart}. The experimental data points are from Refs. 
\protect\cite{Kdat}.
}
\end{figure}

Although our approach relies mainly on experimental 
data, it has some flaws, and the upper bound on $|\kappa_{s}|$
should be considered as order of magnitude estimate only. In particular,
the production of two- and more-meson states should lead to an enhancement 
at the higher energies. On the other hand, the RHS of
eq. (\ref{ungldgh}) is expected to overestimate the GDH integral by far,
because the GDH integrand is not positive definite and 
cancellations can occur. This effect can be
studied by evaluating the GDH integral within hadronic models.
In the case of $\eta$ photoproduction,
phenomenological models \cite{GK95} predict a negative 
contribution to $\kappa_p^2$ of about the same magnitude as our bound. 
This result can be understood by means of eq. (\ref{dd2}). 
Photoproduction of $\eta$ mesons is dominated by the $S_{11}$ resonance
which has a $50\%$ branching ratio to $\eta N$,
and the corresponding $E_{0+}$ multipole enters with a negative 
sign in eq. (\ref{dd2}). For the numerically more important 
contribution of $K$ mesons, we evaluate the 
GDH integral using the model of Mart et al. \cite{mart}.
In Fig. \ref{fig2}, we show a comparison of the transverse 
structure function $\sigma_T$ (eq. (\ref{sigt})) 
and the transverse-transverse structure function $\sigma_{TT'}$ 
(eq. (\ref{sigtt})) for the $K^+\Lambda$- and $K^+\Sigma^0$-channels.
It is clearly seen that $\sigma_T$ overestimates $\sigma_{TT'}$ 
considerably. Moreover, the model calculation predicts a negative sign 
for $\sigma_{TT'}$ (Note that $-\sigma_{TT'}$ has been plotted in
Fig. \ref{fig2}). This sign follows from the fact that the model
predicts a dominance of the multipoles $E_{0+}$ and $M_{1-}$
( see eq. (\ref{dd2})). Since there are only few data on kaon
photoproduction off the neutron, we are not able to perform the
isospin decomposition of eq. (\ref{dhg2}) on the basis of experiments, 
\beq
\label{kappavs}
\kappa_{p/n}^2=(\kappa_{I=1}\pm\kappa_{I=0})^2=\kappa_{I=1}^2
\pm 2\,\kappa_{I=1}\kappa_{I=0} + \kappa_{I=0}^2 \; .
\eeq
Obviously, the strange quark contributes only to the isoscalar 
magnetic moment.
Since the GDH integral for $\kappa_{I=1}^2$ is well saturated 
by pion photoproduction \cite{Wo92}, we do not expect substantial
contributions to $\kappa_{I=1}^2$ from reactions involving
$\phi$, $\eta$ and $K$ mesons.
On the other hand, the remaining two terms are 
not well decribed by the pion data and additional contributions are
needed. 
Present estimates for the isoscalar-isovector
interference cannot even describe the sign of this term properly.
With our model \cite{mart}, we are in the position to isolate the mixed 
term by subtracting $\kappa_p^2$ and $\kappa_n^2$.
Integrating eq. (\ref{dhg}) up to $\nu_{max}=2.2\mbox{ GeV}$,
we find $\kappa_p^2(K)=-0.07$ in qualitative agreement with
Table \ref{restab}. However, the value $\kappa_n^2(K)=-0.02$
indicates that kaon photoproduction contains contributions to
both $\kappa_{I=1}^2+ \kappa_{I=0}^2$ and $2\,\kappa_{I=1}\kappa_{I=0}$,
namely $-0.045$ and $-0.025$, respectively.
We conclude, however,  that the contribution of the $K$ agrees with a negative
sign for $\kappa_{I=0}$ and $\kappa_s$.

In the future, it may be expected 
that the results of the Bonn-Mainz GDH collaboration and similar experiments 
at Jefferson Lab will lead to much more stringent bounds \cite{prop,Bu91}.
In this context it will also be interesting to study the
dependence of the GDH and the Burkhardt-Cottingham
(BC) sum rules on the momentum transfer.
The BC sum rule \cite{BC70} relates the integral 
over the longitudinal-transverse structure function $\sigma_{LT'}$ 
with ground state properties of the nucleon \cite{DD95}. 
Both sum rules have never been measured. They involve experiments with 
longitudinally polarized electrons and nucleon polarization in the 
scattering plane, parallel or perpendicular to 
the momentum of the virtual photon, respectively.
An experiment at Jefferson Lab which has been proposed to measure
$\sigma_{TT'}$ and $\sigma_{LT'}$ \cite{Bu91}, will
determine the structure functions $G_1(\nu,Q^2)$ and $G_2(\nu,Q^2)$ 
from polarized deep-inelastic electron nucleon scattering
at a momentum transfer $Q^2 \approx 0\ldots 2 \mbox{ GeV}^2$. 
This will make it possible to evaluate the BC sum rule \cite{So93},
\beqa
\label{bc}
J_2 (Q^2) &=& m^2\;\int_0^\infty d\nu\; G_2(\nu,Q^2)\\
&=& \frac{\mu}{4}\,G_M(Q^2)\left( \mu\, G_M(Q^2)-G_E(Q^2) \right)\; ,
\nonumber\eeqa
where $\mu$ is the total magnetic moment of the nucleon
and $G_M$ $(G_E)$ is its magnetic (electric) form factor.
Since eq. (\ref{bc}) connects the form factors of the nucleon with the 
sum rule as function of $Q^2$, it has the potential to give estimates 
not only on the magnetic moments but also on the radii.
For the leading $Q^2$ behaviour, we obtain
\beq
\label{radii}
J_2 (Q^2) = \frac{\mu\kappa}{4} - Q^2\; \frac{\mu}{24}
\left( (\mu+\kappa)<r_M^2 > - < r_E^2 >\right) + O( Q^4)\nonumber\;\, .
\eeq
As in the case of the GDH sum rule, 
the different reactions add incoherently, thus defining the contributions 
of the individual production channels to the anomalous magnetic moment and 
a particular combination of electric and magnetic charge radii.

We conclude that detailed investigations of the strangeness contributions to 
the GDH and BC sum rules could give interesting information about strange 
quark effects in the nucleon. The results would be complementary to the 
considerable experimental activities at MAMI, MIT/Bates and Jefferson
Lab to measure the contribution of strangeness by parity violating 
electron scattering.

\section*{Acknowledgement}
This work has been supported by the Deutsche Forschungsgemeinschaft
(SFB 201). HWH has also been supported by the German Academic Exchange 
Service (Doktorandenstipendium HSP II/AUFE).
We would like to thank G. Kn{\"o}chlein for providing us
with the results of his model calculations for the GDH integral
and M.J. Musolf for some useful comments.

\newpage
\begin{center}
{\Large\bf Figures}
\end{center}
\medskip
\medskip

\medskip

\end{document}